\documentclass{aa}  
\usepackage{graphicx}
\usepackage{txfonts}
\usepackage{natbib}
\bibpunct{(}{)}{;}{a}{}{,}
\begin{document}
%
   %\title{}
   \title{An IPHAS-based search for accreting very low-mass objects using VO tools}
   \author{L. Valdivielso
          \inst{1}
          \and
          E.~L. Mart\'\i n
          \inst{1,}\inst{2}
	  \and
	  H. Bouy
	  \inst{1,}\inst{3}
	  \and
	  E. Solano
	  \inst{4,}\inst{5}
	  \and
          J.~E. Drew
          \inst{6,}\inst{7}
	  \and
          R. Greimel
          \inst{8,}\inst{9}
	  \and
	  R. Guti\'errez
	  \inst{4,}\inst{5}
          \and
          Unruh, Y.~C.
          \inst{10}
          \and
          Vink, J.~S.
          \inst{11}}

   \offprints{L. Valdivielso (\email{lval@iac.es})}

   \institute{$^1$ Instituto de Astrof\'\i sica de Canarias, C/ V\'\i a L\'actea, s/n, E-38200 - La Laguna, Tenerife, Spain\\
         \email{lval@iac.es,ege@iac.es,bouy@iac.es}\\
	$^2$ University of Central Florida, Dept. of Physics, PO Box 162385, Orlando, FL32816-2385, USA \\ 
	$^3$ Astronomy Department, University of California, Berkeley, CA 94720, USA\\
  	 $^4$ Laboratorio de Astrof\'\i sica Espacial y F\'\i sica Fundamental (LAEFF-INTA), Apdo.78, E28691 Villanueva de la Ca\~nada, Madrid, Spain\\
        \email{esm@laeff.inta.es,raul@laeff.inta.es}\\
         $^5$ Spanish Virtual Observatory Thematic Network\\   
         $^6$ Imperial College of Science, Technology and Medicine,Blackett Laboratory, Exhibition Road, London,  SW7 2AZ, U.K.\\
         \email{j.drew@imperial.ac.uk}\\ 
         $^7$Centre for Astrophysics Research, University of Hertfordshire, College Lane, Hatfield AL10 9AB, U.K.\\
	 $^8$Institut für Physik, Universität Graz, Universitätsplatz 5, A-8010 Graz, Austria \\
        \email{rgreimel@gmail.com}\\
         $^9$Isaac Newton Group of Telescopes, Apartado de correos 321,E-38700 Santa Cruz de la Palma, Tenerife, Spain \\
         $^{10}$Astrophysics Group, Blackett Laboratory,  Imperial College London, SW7 2AZ\\
        \email{y.unruh@imperial.ac.uk}\\
         $^{11}$Armagh Observatory, College Hill, Armagh BT61 9DG, Northern Ireland, UK\\
        \email{jsv@arm.ac.uk}
   	}
   \date{Received ...; accepted 15 Oct 2008}

% \abstract{}{}{}{}{} 
% 5 {} token are mandatory
 
  \abstract
   { }
   {The main goal of this paper is to prove that accreting very low-mass stars and brown dwarfs can be identified in IPHAS, a H$\alpha$ emission survey of the northern Milky Way.Full exploitation of the IPHAS database and a future extension of it in the southern hemisphere will be useful to identify very low-mass accreting objects near and far well-known star forming regions.}
   {We have used Virtual Observatory tools to cross-match the IPHAS catalogue with the 2MASS catalogue. We defined photometric criteria to identify H$\alpha$ emission sources with near-infrared colours similar to known young very low-mass stars and brown dwarfs. 4000 candidates were identified that met our criteria over an area of 1600 square degrees. 
%* these numbers should be checked 
We present low-resolution optical spectra of 113 candidates. Spectral types have been derived for the 33 candidates that have spectroscopically confirmed H$\alpha$ emission, negligible reddening and M spectral class. We have also measured H$\alpha$ emission and investigated the NaI doublet (818.3 nm, 819.5 nm) in these 33 objects.}
   {We confirm that 33  IPHAS candidates have strong H$\alpha$ indicative of disk accretion for their spectral type. 23 of them have spectral class M4 or later, of which 10 have classes in the range M5.5--M7.0 and thus could be very young brown dwarfs.  Also many objects have weak NaI doublet, an indication of low surface gravity.}
   {We conclude that IPHAS provides a very valuable database to identify accreting very low-mass stars and brown dwarfs, and that Virtual Observatory tools provide an efficient method for identifying these objects over large areas of the sky. Based on our success rate of 23 H$\alpha$ emission objects with spectral type in the range M4--M7 out of 113 candidates with spectroscopic follow-up, we estimate that there could be hundreds of such objects in the full IPHAS survey. 
%* could we make a more accurate estimate ? 
}

    \authorrunning{L. Valdivielso et al.}
    \titlerunning{VLM objects search on IPHAS}

   \maketitle
%
%________________________________________________________________

\section{Introduction}
%bd's, VO

Since the first unambiguous discovery of brown dwarfs (BDs) \citep{1995Natur.377..129R, 1995Natur.378..463N} this field has progressed rapidly. Considerable observational effort has been devoted to identifying BDs in the known nearby star-forming regions (SFRs) and young open clusters. 
These objects provide crucial information on the dependence of key physical properties, such as disk properties, multiplicity and the shape of the Initial Mass Function on primary mass.

Due to mass accretion processes, many young low-mass stars and BDs show H$\alpha$ emission stronger than the emission expected from chromospheric activity. 
Studying the H$\alpha$ equivalent width and the spectral type using low-resolution spectra, it can be determined whether or not a star is accreting \citep {2003AJ....126.2997B}. Thus H$\alpha$ surveys have the potential to identify very young stars and BDs that are still accreting from their disks. 

Some H$\alpha$ searches for very low-mass (VLM) stars and BDs have already been instrumental in detecting these objects, such as for example those carried out in high latitude molecular clouds \citep{1996A&AS..116..467M} or in Orion OB1 \citep{2001Sci...291...93B}. Evidence of accretion disks and infrared excess in young VLM stars and BDs has been confirmed in several SFRs, such as $\rho$ Ophiuchus \citep{1999AJ....117..469W}, the Trapezium \citep{2001ApJ...558L..51M}, Chamaleon I \citep{2001A&A...376L..22N}, IC348 \citep{2003ApJ...592..282J} or $\sigma$ Orionis \citep{2002A&A...382L..22O, 2002A&A...384..937Z, 2003ApJ...592..266M}. Some spectroscopic studies have emphasised on the analysis of the H$\alpha$ emission of substellar objects and led to the detection of very cold objects such as S Ori 55 \citep {2002ApJ...569L..99Z} or S Ori 71 \citep{2002A&A...393L..85B}.

The INT Photometric H$\alpha$ survey of the Northern Galactic Plane (IPHAS) is a valuable source for discovering young VLM stars and BDs using H$\alpha$ emission. 
It covers 1800 square degrees of the northern Milky Way in the latitude range $-$5\degr $<$ b $<$ 5\degr. It provides (Sloan) r', i' and narrowband H$\alpha$ photometry down to a magnitude limit of r'$\sim$ = 20 (10 $\sigma$). The data is taken using the Wide Field Camera (WFC) on the 2.5m Isaac Newton Telescope (INT) \citep{2005MNRAS.362..753D, 2007arXiv0712.0384G}.

So far the overwhelming majority of the surveys for young VLM objects are concentrated in the known SFRs and nearby young clusters.Take as an example recent research in Taurus \citep{2006A&A...446..485G}, Orion OB1A and OB1B \citep{2006RMxAC..26...37D}, Chamaeleon I \citep{2004ApJ...602..816L}, Chamaeleon II and Ophiuchus \citep{2007ApJ...657..511A} or $\sigma$Orionis \citep{2007A&A...470..903C}.

The IPHAS survey offers a complementary approach because it allows us to use H$\alpha$ as a primary selection criterion, and it provides a wide area coverage around or outside the well-known SFRs and clusters.\\

The Virtual Observatory (VO) is a recent initiative with the goal of managing large databases in an organised manner in order to make an efficient use of astronomical archives. In
this work VO tools have been applied for the first time to the search for young VLM objects. We have made a cross-correlation of an early version of the IPHAS point source catalogue with the Two Micron All Sky Survey (2MASS) point source catalogue \citep{2006AJ....131.1163S}. The rest of the paper is organised as follows: Section 2 describes the selection criteria used for identification of young VLM candidates. Section 3 deals with spectroscopic follow-up observations of 113 candidates selected using our criteria. Section 4 discusses the analysis of the spectra leading to the determination of spectral types and equivalent widths. Section 5 presents the main results of our study.

\section{Sample selection}
The VO \ offers the possibility of efficiently cross correlating large multi-wavelength databases. \emph{Aladin} \footnote{http://aladin.u-strasbg.fr/aladin.gml.} is an interactive software sky atlas allowing the user to visualise digitised astronomical images, superimpose entries from astronomical catalogues or databases, and interactively access related data and information from the Simbad database, the VizieR service and other archives for all known sources in the field.
 
We have used \emph{Aladin} to look for new VLM objects via cross-correlation of the IPHAS catalogue with 2MASS. The photometric data available in these catalogues provides magnitudes in H$\alpha$, r', i', J, H and K for the selected objects. \\ 
We used an early version of the IPHAS point source catalogue that covered about 1200 square degrees. Our search was restricted to the RA range from 18 to 05 hours. We identified 4000 candidates that met the following criteria: 

\begin{itemize} 
\item IPHAS-2MASS coincidence in coordinates within 1\arcsec \ . 
\item The IPHAS sources should  be classified as stellar or probably stellar in the r', i' and H$\alpha$ bands and have colours in the range 1.1 $<$ r'$-$H$\alpha$ $<$ 3.0 (to avoid potential artifacts and select objects with H$\alpha$ in emission according to \citet{2005MNRAS.362..753D} who show that M6 dwarfs have IPHAS colours r'$-$H$\alpha$ $=$ 1.06 for  E(B$-$V) $=$ 0.0, and r$-$H$\alpha$ $=$ 1.14 for  E(B$-$V) $=$ 1.0); and magnitudes i' $<$ 18.5.
\item The 2MASS sources should have \emph{qflag} A or B; colours in the range 0.7 $<$ J$-$H $<$ 1.3 (to discard strong reddened objects and red giants); and 0.4 $<$ H$-$K $<$ 1.1 (for selecting cool photospheres and/or infrared excess). 
\end{itemize}  

\section{Follow-up spectroscopic observations}
We have carried out 3 campaigns of follow-up low-resolution spectroscopic observation of our targets. Altogether, spectra for 113 candidates have been obtained, which represents only 3\% of our total sample. We now discuss these runs in chronological order: 
 
\subsection{William Herschel Telescope observations}
Our first run took place on August 1--2, 2006 with the ISIS long slit spectrograph on the 4.2m William Herschel Telescope in La Palma as part of the international time program led by Janet Drew. The R158R grating in red arm was used. The instrumental setup gave a dispersion of 1.63 \AA/pixel with a wavelength range of 5400--10300\AA . We took spectra with exposure times 
ranging from 700 to 1800s for 35 young VLM candidates.
Due to poor weather conditions in the first night (variable seeing of 2--3\arcsec) the slit width was 2.0\arcsec (FWHM $\sim$ 15\AA). In the second night we had much better seeing, ranging from 0.7\arcsec to 1\arcsec, and hence a slit width of 1.0\arcsec was used (FWHM $\sim$ 7\AA). Each night two flux calibration  standard stars were observed, and Ne-Ar arcs and lamp flat-fields were taken at the beginning of the nights.
Bias and flat-field correction has been applied to all our CCD frames. 
Wavelength calibration and instrument response corrections were made on the science spectra using standard tasks in the IRAF environment.  
 \onltab{1}{
\begin{table*}
\caption{Spectral indices obtained for the IPHAS candidates.}             
\label{tabla3}      
\centering          
\begin{tabular}{c c c c c}     % 4 columns 
\hline\hline 
 Object& PC3 & PC6& TiO1& TiO2  \\
\hline
J001649.57+654241.8   &  1.20 &    9.50&      1.52 &	 1.22	 \\
J001655.91+654732.8   &  1.23 &    8.08&      1.55 &	 1.19  \\
		      &  1.16 &    7.68&      1.44 &	 1.10  \\
J011443.00+620820.9   &  1.19 &    6.47&      1.15 &	 1.02	 \\
J012348.67+614931.8   &  1.13 &    4.01&      1.21 &	 1.05	  \\
J013720.01+645957.7   &  1.11 &    4.14&      1.30 &	 1.12	  \\
J023616.00+615609.6   &  0.99 &    2.54&      1.17 &	 0.94	  \\
J035449.17+530903.3   &  1.06 &    3.59&      1.16 &	 1.03	  \\
J035823.95+522312.6   &  1.00 &    2.01&      1.03 &	 0.88	  \\
J042450.68+455330.2   &  1.21 &    3.48&      1.14 &	 0.93	\\
J183034.15+003800.6   &  1.35 &    7.46&      1.22 &	 0.92	 \\
J183753.25+001849.2   &  1.35 &    5.77&      1.32 &	 1.17	\\
J192656.04+211438.0   &  1.05 &    1.76&      1.03 &	 0.96	 \\
J202050.40+394243.6   &  1.11 &    2.42&      0.99 &	 0.98	  \\
J202434.30+422126.4   &  1.06 &    6.91&      1.66 &	 1.29  \\
J202437.28+385806.9   &  1.12 &    4.74&      1.37 &	 1.18	 \\
J202455.53+424504.0   &  1.12 &    4.94&      1.56 &	 1.31  \\
J202759.82+390418.2   &  1.12 &    3.00&      1.06 &	 0.95	 \\
J204218.54+395723.8   &  1.00 &    2.77&      1.04 &	 0.99	\\
J204350.68+400108.8   &  0.92 &    1.78&      1.05 &	 1.00	 \\
J204704.82+434911.4   &  1.08 &    2.12&      1.05 &	 0.97	  \\
J205613.08+443424.2   &  1.28 &    7.10&      1.28 &	 1.36	  \\
J205701.63+434138.7   &  1.28 &    8.90&      1.75 &	 1.46	\\
J205702.69+434143.7   &  1.09 &    7.43&      1.72 &	 1.33	\\
J210404.87+535124.4   &  1.43 &    7.95&      1.26 &	 1.14	 \\
   		      &  1.17 &    7.09&      1.45 &	 1.22	\\
   		      &  1.30 &    7.59&      1.47 &	 1.27	  \\
J213528.41+575823.0   &  1.10 &    3.58&      1.15 &	 1.05	  \\
J213545.87+573640.1   &  1.09 &    2.97&      1.23 &	 1.01	  \\
J213938.83+575451.4   &  1.09 &    4.01&      1.23 &	 1.03	\\
		      &  1.13 &    3.62&      1.23 &	 1.06	  \\
J214547.73+564845.7   &  1.29 &    6.43&      1.42 &	 1.22  \\
J214625.99+572829.0   &  1.05 &    3.24&      1.14 &	 1.08	 \\
		      &  1.18 &    5.15&      1.27 &	 1.20	  \\
J222025.22+605423.7   &  1.11 &    3.11&      1.24 &	 1.03	  \\
J224830.78+611417.9   &  1.15 &    5.36&      1.50 &	 1.18	   \\
		      &  1.14 &    5.70&      1.45 &	 1.13	  \\
J225331.29+623543.3   &  1.26 &    8.62&      1.36 &	 1.06	 \\
J225923.71+614138.6   &  1.23 &    4.75&      1.05 &	 0.91	  \\
\hline                  
\end{tabular}
\end{table*}
}

\subsection{Northern Optical Telescope observations}
Our second run occurred on October 10--14, 2006 with the ALFOSC spectrograph on the 2.5m Nordic Optical Telescope in La Palma. \emph{Grism 5} was used corresponding to a dispersion of 3.1 \AA/pixel and a wavelength range of 5000--10250\AA. A slit width of 1.0\arcsec (FWHM $\sim$ 16\AA) was used and exposure times ranged from 300 s to 2400 s.  56 new candidates were observed. One flat-field and one arc lamp exposure were taken immediately before or after each science target. A flux calibration standard star was observed at the beginning of the nights. 
Bias correction was performed using an average bias for each night, and flat-field correction was made with the normalised flat-field exposure obtained at the same telescope position as the scientific spectrum. This procedure improved the correction of the fringing pattern in the CCD detector response over the correction that 
was obtained using an average flat-field.  
The calibration in wavelength was done using the images of He-Ne-Th-Ar arcs obtained after each scientific image. Tests were done calibrating with the emission lines of sky that appear in the spectra instead of with the arcs, nevertheless the results of the calibration were not better. 
Tests were also made using different polynomial functions to fit the wavelength solution. Finally, a 4th order Legendre polynomial was chosen. We measured that the displacements of known sky airglow emission lines with respect to the expected positions are generally less than $\sim$2\AA \ , which is acceptable because it is less than 1/8 of our spectral resolution. The standard HD227900 was used for flux calibrations. All data reductions were carried out using IRAF.
%(http://www.not.iac.es/instruments/alfosc/)

\subsection{Lick Observatory Shane Telescope observations}

In July 2007 we used the KAST spectrograph on the 3m Shane Telescope
at Lick observatory with the 600/7500 (2.32 \AA/pixel) and
300/7500(4.6 \AA/pixel) gratings in the red arm only with a 2\arcsec \ slit
width and exposure times from 300s to 2400s for a total of 25 candidates. The
wavelength range covered was of 5600--8500 \AA \ for the 600/7500 grating and
5000-10500 \AA \ for the 300/7500 grating. Ne arcs and flat-field lamp exposures were taken for each position of the scientific images. Part of the observations of the last night were affected by adverse weather (cirrus). In general the quality of the data collected during this campaign is 
poorer than the two previous runs. During the first two nights the 600/7500 grating was used 
(FWHM $\sim$ 5\AA). Although the weather was good, with seeing of $\sim$1.5\arcsec, the spectra were very noisy and weak, sometimes so much that it was not possible to extract them. 
For the 2 last nights we decided to change to the 300/7500 grating (FWHM $\sim$ 9\AA) 
in order to improve the signal to noise ratio of the spectra.  
For the reduction, flat-field normalised images obtained after each scientific exposure were applied. For wavelength calibration Ne arcs also taken after each exposure was used. Depending on the night and the grating, different spectrophotometric standards were used for flux calibrations. All data reductions were carried out using IRAF. 
 
\section{Spectral analysis}

We found that 42 of our 113 targets have strong  H$\alpha$ emission. 
Therefore we have had a 37\% success rate in the confirmation of H$\alpha$ emission 
among our candidates.We find 12 of the 42 confirmed targets in the IPHAS catalogue of emission-line sources of \citet{2008MNRAS.384.1277W}. For the sample selection we used a preliminary version of the IPHAS catalogue which was available to us in July 2006. Due to the improvement of the H$\alpha$ zero point calibration on the IPHAS catalogue, although all of our candidates where selected as likely H$\alpha$ emitters with r'$-$H$\alpha$ $>$ 1.1, most of the non H$\alpha$ emitters are out of the main stellar locus of emission-line sources of \citet{2008MNRAS.384.1277W} in the IPHAS colour-colour plane, and only 48 of them fit the selection criteria according to the new photometric IPHAS catalogue. According to this, we have 48 observed candidates that really fit the criteria and from those 40 of them show H$\alpha$ emission which is 83\% success rate. In the following analysis we only retain the objects with confirmed H$\alpha$ emission.

\subsection{Spectral type determination}

The spectral type has been determined using 3 different methods: 
(a) the \emph{Hammer} code; (b) $\chi^2$ fitting with template spectra; and (c) 
measurement of spectral indices that have a known relation to spectral classification. 

\subsubsection{Comparison with template spectra: The \emph{Hammer} code}
\emph{Hammer} \footnote{See http://www.cfa.harvard.edu/~kcovey/thehammer.} is a tool written in IDL for the analysis of stellar spectra \citep{2007AJ....134.2398C}. With this routine  each spectrum is analysed automatically to predict the spectral type by calculating a list of spectral indices (measuring the strength of CaH and TiO bands) to a library, considering the uncertainty of each index and also do  measurements of H$\alpha$ as tracer of activity in late type stars. In addition, it allows the user to manually compare and to change the final spectral type that is initially assigned automatically by the program via a rough comparison with the sample of reference spectra. 
The routine has been applied to all our targets with confirmed emission in H$\alpha$. In most cases, it was necessary to select interactively the spectral type. Nevertheless, applying this method is useful because it allows to quickly inspect different comparisons of the targets with templates and to derive a first estimate of the spectral type.
\begin{figure}
   \centering
   \resizebox{\hsize}{!}{\includegraphics[width=\textwidth]{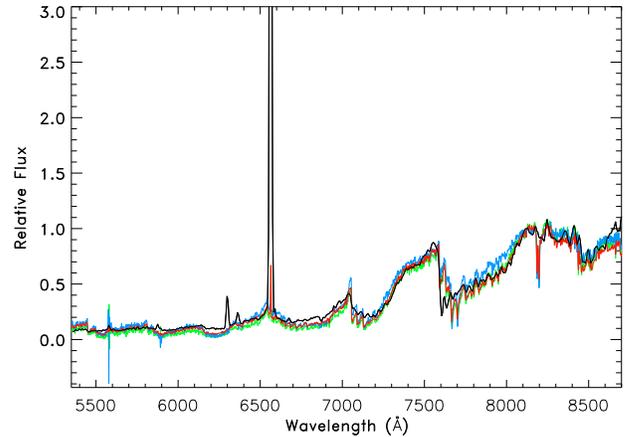}}
      \caption{An example of spectral type determination using the $\chi$$^2$ method. The best fit (M6) is represented in red colour overlaid on the observed spectrum J210404.87+535124.4 (black). For comparison we represent also the M5 (blue) and M7 (green) reference objects. The spectra are normalised at 8123\AA.
              }
         \label{fig1}
   \end{figure}

\subsubsection{$\chi$$^2$ method}
An IDL program has been written to calculate the $\chi$$^2$ value comparing our target spectra with those of M-type reference field dwarfs. 
We found the best spectral type by minimisation of the $\chi$$^2$ value. 
As reference for the spectral classification, stars of the IAC ultracool dwarf catalogue were used \citep{2005AN....326.1026M}. We selected one dwarf for each spectral subclass between M0 and M9. The dwarfs selected were: SDSS J125350.99-001010.3 (M0), SDSS J004830.11-003029.1 (M1), SDSS J000201.55-010636.5 (M2), SDSS J011113.97-003128.4 (M3), SDSS J002328.15+001709.02 (M4), SDSS J113922.08+000048.8 (M5), SDSS J150026.34-003928.0 (M6), SDSS J100218.61-000408.46 (M7), SDSS J135859.02-005357.5 (M8), and SDSS J024958.36-003409.99 (M9). The spectra of the catalogue 
were resampled using the task \emph{linterp} and normalised in the wavelength range of $\lambda$=8110--8136 \AA \ (central wavelength $\lambda$=8123 \AA) to make the calculation of $\chi$$^2$.\\

In Fig. \ref{fig1} we display an example of the application of the $\chi$$^2$ method for the spectral type determination in one of our targets. The object spectrum (black line) is compared with the template that yields the minimum $\chi$$^2$ value (red line), and with the templates that have adjacent earlier or later spectral subclass (blue and green lines, respectively).  
In general the method works fine, although there were some objects that did not fit properly to any spectral template. This method did not provide trustworthy results for noisy or highly reddened spectra because the calculation of $\chi$$^2$ is based on the difference point to point between two spectra. 
\begin{table*}
\caption{Spectroscopic results.}             
\label{table:1}      
\centering          
\begin{tabular}{c c c c l l l l}     % 8 columns 
\hline\hline       
  IPHAS name/position & SpT& SpT& SpT&  SpT & W(H$\alpha$)  \\ % &    \\
 J[RA(2000)+Dec.(2000)] & (\emph{Hammer})& ($\chi$$^2$)& (indices)&  (adopted)& (\emph{Hammer})& W(H$\alpha$) \\ % &  W(Na I)  \\                      
\hline                    
J001649.57+654241.8	& M5	&    M7 &	M4.5	& M5.5$\pm$1.0 &  140.3 & 250$\pm$80  \\ % & $<$ 2.9$\pm$0.3   \\
J001655.91+654732.8	& M6	&    M7 &	M5.5	& M6$\pm$0.5   &   57.7 &  74$\pm$15  \\ % & $<$ 2.0$\pm$1.9   \\
        		& M5	&    M7 &       M5.5    & M6$\pm$1.0   &   96.1 & 133$\pm$15  \\ % &		      \\
J011443.00+620820.9	& M4	&    M5 &	M4.5	& M4.5$\pm$0.5 &  192.7 & 276$\pm$50  \\ % & $<$  1.9$\pm$0.2  \\
J012348.67+614931.8	& M4	&    M4 &	M4.5	& M4$\pm$0.5   &  135.6 & 230$\pm$30  \\ % & $<$  3.0$\pm$0.3  \\
J013720.01+645957.7	& M4	&    M4 &	M4.5	& M4$\pm$0.5   &  112.7 & 170$\pm$30  \\ % & $<$  3.3$\pm$0.3  \\
J023616.00+615609.6	& M1	&    M4 &	-	& M2.5$\pm$1.0 &  133.7 & 205$\pm$30  \\ % & $<$  3.0$\pm$0.3  \\
J035449.17+530903.3	& M3	&    M4 &	M2.5	& M3$\pm$0.5   &   93.7 & 164$\pm$30  \\ % & $<$  1.7$\pm$0.2  \\
J035823.95+522312.6	& M1	&    M2 &	-	& M1.5$\pm$0.5 &  147.2 & 214$\pm$30  \\ % &   - 	      \\
J042450.68+455330.2	& M3	&    M4 &	-	& M3.5$\pm$0.5 &   21.6 & 122$\pm$30  \\ % &   - 	      \\
J183034.15+003800.6	& M6	&    M8 &	M5.5	& M6.5$\pm$1.0 &  185.8 & 161$\pm$50  \\ % & $<$ 1.8$\pm$0.2   \\
J183753.25+001849.2	& M6	&    M7 &	M5	& M6$\pm$1.0   &   16.3 &  94$\pm$30  \\ % & $<$ 1.5$\pm$0.2   \\
J192656.04+211438.0	& M0	&    M0 &	M$<$2.5 & M0$\pm$0.5   &   35.6 &  52$\pm$5   \\ % &   - 	      \\
J202050.40+394243.6	& M2	&    M2 &	M3	& M2$\pm$0.5   &   83.5 & 140$\pm$15  \\ % &   - 	      \\
J202434.30+422126.4	& M5	&    M6 &	M5.5	& M5.5$\pm$0.5 &   23.5 &  32$\pm$3   \\ % & $<$ 1.7$\pm$0.8   \\
J202437.28+385806.9	& M5	&    M6 &	M4.5	& M5$\pm$0.5   &   93.2 & 138$\pm$30  \\ % & $<$ 1.3$\pm$0.2   \\
J202455.53+424504.0	& M5	&    M5 &	M5	& M5$\pm$0.5   &   88.6 & 127$\pm$30  \\ % & $<$ 3.0$\pm$0.2   \\
J202759.82+390418.2	& M3	&    M4 &	-	& M3.5$\pm$0.5 &  106.2 & 143$\pm$10  \\ % &   - 	      \\
J204218.54+395723.8	& M2	&    M2 &	-	& M2$\pm$0.5   &   14.5 &  19$\pm$2   \\ % &   - 	      \\
J204350.68+400108.8	& M1	&    M2 &	-	& M1.5$\pm$0.5 &   44.2 &  71$\pm$8   \\ % &   - 	      \\
J204704.82+434911.4	& M1	&    M2 &	M2.5	& M2$\pm$0.5   &  181.0 & 245$\pm$50  \\ % &   - 	      \\
J205613.08+443424.2	& M6	&    M6 &	M5	& M6$\pm$0.5   &  210.2 & 261$\pm$80  \\ % & $<$ 3.0$\pm$0.5   \\
J205701.63+434138.7	& M7	&    M8 &	M5.5	& M7$\pm$1.0   &   40.8 &48.5$\pm$20  \\ % & $<$ 1.0$\pm$0.1   \\
J205702.69+434143.7	& M5	&    M6 &	M5.5	& M5.5$\pm$0.5 &  117.3 & 251$\pm$50  \\ % & $<$ 2.2$\pm$0.3   \\
J210404.87+535124.4	& M6	&    M6 &	M6	& M6$\pm$0.5   &  282.4 & 245$\pm$90  \\ % & $<$ 0.4$\pm$0.2   \\
			& M6	&    M6 &       M5.5    & M6$\pm$0.5   &  288.6 & 580$\pm$100 \\ % &		      \\
			& M6	&    M7 &       M5.5    & M6$\pm$0.5   &  456.1 & 570$\pm$200 \\ % &		      \\
J213528.41+575823.0	& M4	&    M4 &	M4.5	& M4$\pm$0.5   &  126.2 & 202$\pm$40  \\ % & $<$ 3.8$\pm$0.3   \\
J213545.87+573640.1	& M3	&    M4 &	M4	& M4$\pm$0.5   &  189.8 & 221$\pm$80  \\ % & $<$ 3.7$\pm$0.3   \\
J213938.83+575451.4	& M4	&    M4 &	M5	& M4.5$\pm$0.5 &  105.4 & 198$\pm$30  \\ % & $<$ 2.6$\pm$0.4   \\
        		& M4	&    M4 &       M4.5    & M4$\pm$0.5   &  124.0 & 166$\pm$40  \\ % &		      \\
J214547.73+564845.7	& M4	&    M6 &	M5	& M5$\pm$1.0   &  147.8 & 131$\pm$55  \\ % & $<$ 1.3$\pm$0.3   \\
J214625.99+572829.0	& M2	&    M3 &       M2.5    & M2.5$\pm$0.5 &  243.8 & 360$\pm$60  \\ % &   - 	      \\
        		& M4	&    M5 &       M4.5    & M4.5$\pm$0.5 &  388.3 & 541$\pm$180 \\ % &		      \\
J222025.22+605423.7	& M4	&    M4 &	M4.5	& M4$\pm$0.5   &  233.8 & 370$\pm$100 \\ % & $<$ 2.6$\pm$0.4   \\
J224830.78+611417.9	& M5	&    M5 &	M4.5	& M5$\pm$0.5   &  137.7 & 204$\pm$70  \\ % & $<$ 2.8$\pm$1.0   \\
        		& M5	&    M5 &       M5      & M5$\pm$0.5   &  143.7 & 190$\pm$30  \\ % &		      \\
J225331.29+623543.3	& M6	&    M7 &	M5	& M6$\pm$1.0   &  194.6 & 350$\pm$100 \\ % &   - 	      \\
J225923.71+614138.6	& M5	&    M5 &       M5      & M5$\pm$0.5   &  250.1 & 540$\pm$50  \\ % & $<$ 2.5$\pm$0.3   \\
\hline                  
\end{tabular}
\end{table*}

\subsection{Equivalent widths}
\label{ew}
 
\subsubsection{Spectral indices}
The spectral indices PC3, PC6, TiO1 and TiO2 of the objects with emission in H$\alpha$ have been determined using the IRAF task \emph{sbands} and the results (see Table \ref{tabla3} only available in electronic form) have been compared with those of \citet{1999AJ....118.2466M} to determine the spectral type of our candidates. 
For 7 objects we could not find a reasonable agreement between the different indices (within 2 spectral subclasses), and hence this method was not used for them. 

\emph{Hammer} provides measurements of the equivalent width of H$\alpha$ simultaneously with the spectral type determination. However, it does not provide reliable errors for our spectra  because \emph{Hammer} calculates them by propagating the error estimates provided by the SDSS spectral reduction pipeline.  

We derived the H$\alpha$ emission equivalent width of our targets by direct integration of the line profile using the IRAF task \emph{splot}. The uncertainty was estimated via measurements performed for extreme choices of continuum level and line integration wavelength range made by visual inspection of the spectra. 

 We have tried to measure the NaI doublet equivalent width at 818.3 nm and 819.5 nm on all of our objects as low surface gravity indicator, but this part of the spectrum is affected by residual fringing and we can not give a reliable estimation of the equivalent width. To investigate the surface gravity and for comparison purposes, simulations of how the residual fringing affects the doublet on the reference objects have been done. We have degraded the resolution of the reference objects to the resolution of our targets and introduced different residual fringing according to the residuals in our objects. For statistical and  comparison purposes with our targets also measurements of  NaI doublet index with defined blue, central and red bandpasses (8130-8171\AA, 8172-8207\AA, 8235-8265\AA \ respectively due to low resolution of the spectra and visual inspection) have been used for quantitative estimation of the variation with the residual fringing on the objects. The results are shown and discussed in section \ref{secsodio}.

\section{Results}

\begin{figure}
   \centering

  \resizebox{\hsize}{!}{\includegraphics[width=\textwidth]{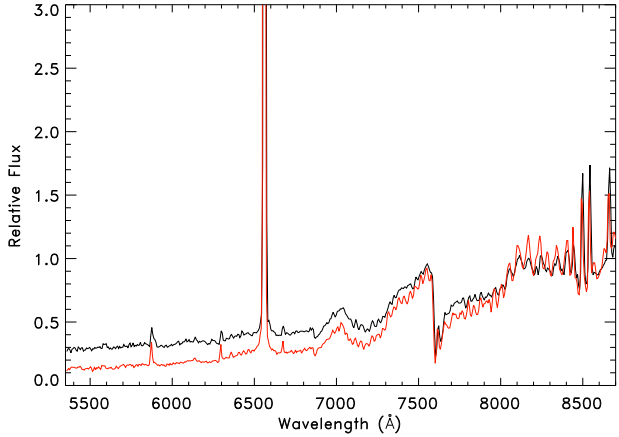}}      
\caption{Comparison of the spectra of IPHAS J214625.99+572829.0 observed on two different nights. 
There is significant variability in the continuum and the H$\alpha$ emission. The spectra are normalised at 8123\AA. 
              }
         \label{fig2}
\end{figure}

\subsection{Definition of the sample of M-type H$\alpha$ emitters}
9 of our 42 targets with confirmed H$\alpha$ emission have very red continuum slopes on the red part of the spectrum, indicative of high reddening and are classified as early-M or late-K objects. They do not have late-M spectral type and thus are not considered in the rest of this paper. Our results are focused on the 33 M-type H$\alpha$ emission objects that have negligible reddening according to our visual inspection of the comparison with reference dwarfs of known spectral types.  

In Table \ref{table:1} we give our results obtained from the 3 different methods of  determination from the spectral type (SpT) and the measurements of equivalent widths 
(\emph{W}) of H$\alpha$ (using Hammer and IRAF) and the Na I doublet. \\
6 of our targets were observed more than once to check for variability. 
They have more than one data line in Table \ref{table:1}. 
We find that observations of the same objects obtained at different epochs and with different instruments give consistent spectral types within the error bars except for object IPHAS J214625.99+572829.0 which shows significant variability (Fig.\ref{fig2},\ref{fig4}).This object also displays strong and variable H$\alpha$ emission. Two more objects (IPHAS J210404.87+535124.4, J001655.91+654732.8) show H$\alpha$ variability larger than the error bars but no significant change in spectral type. The three remaining targets observed more than once do not have H$\alpha$ variability larger than the error bars. These examples indicate that the observed H$\alpha$ emission is not due to occasional flares, because the strong H$\alpha$ emission is observed at several epochs. Emission line variability is an additional indicator for youth and has been found previously for young, accreting brown dwarfs  \citep{2006ApJ...638.1056S}. The variability might be due to co-rotating hot spots, changes in the accretion flow geometry or changes in the mass accretion rate onto the central object \citep{1995A&A...299...89B}.\\

\begin{figure}
\includegraphics[width=0.33\textwidth,angle=270]{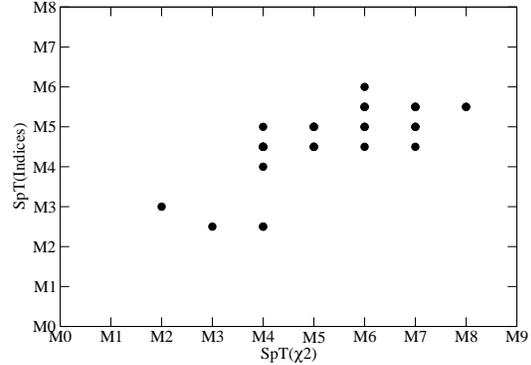}
\includegraphics[width=0.33\textwidth,angle=270]{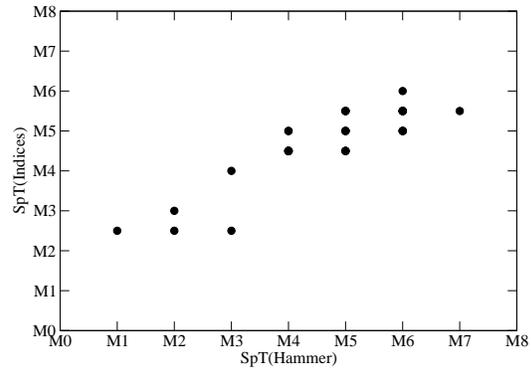}	
\includegraphics[width=0.33\textwidth,angle=270]{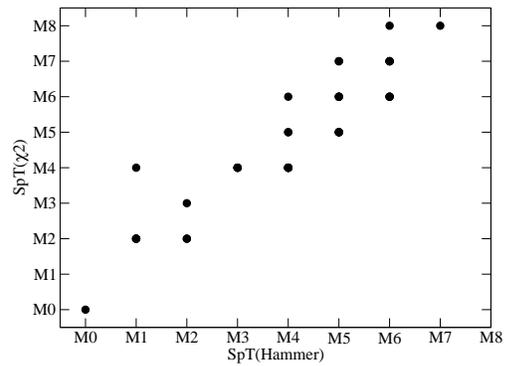}
\caption{Comparison of spectral type determinations using different methods.}
\label{fig3}
\end{figure}
In Fig. \ref{fig3} the results of the three different methods of spectral type determination are compared. \emph{Hammer} tends to give earlier types than the $\chi$$^2$ method. 
The discrepancies between these two methods is in general within one spectral subclass, but there are a few examples of larger differences. The spectral types obtained from the spectral indices are sometimes intermediate between \emph{Hammer} and the $\chi$$^2$ method, but for the latest subclasses in our sample they tend to agree more with \emph{Hammer} than with the $\chi$$^2$. 
We have adopted spectral subclasses by computing the average of the 3 methods, and we have 
rounded to the nearest subclass in steps of 0.5 subclasses.\\

With respect to the measurements of equivalent width of H$\alpha$ it is observed that the values derived with \emph{Hammer} are systematically lower than the ones obtained by manual integration of the line profile. We consider that the latter ones are more reliable than the \emph{Hammer} results because they rely on careful visual inspection of the continuum and line profile on each spectrum. Hence, we adopt the manual values of H$\alpha$ equivalent widths and error bars for the subsequent discussion. \\

\subsection{Disk accretion}

\begin{figure}
   \centering
   \resizebox{\hsize}{!}{\includegraphics[width=\textwidth, angle=270]{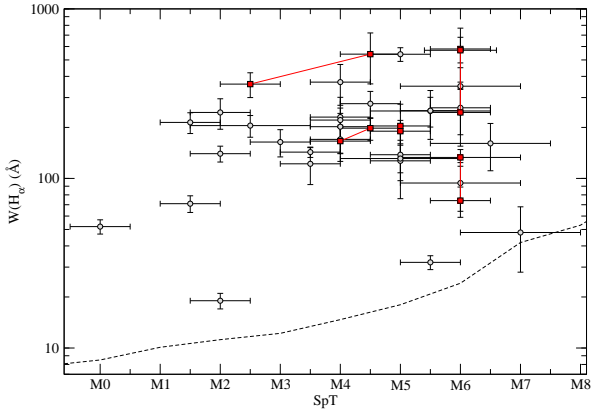}}
      \caption{H$\alpha$ equivalent width against adopted spectral type for our objects. 
The dashed line denotes the dividing line between chromospheric activity and disk accretion. 
Our objects are clearly above the dashed line and hence they are likely undergoing mass accretion.
              }
         \label{fig4}
   \end{figure}

In order to discern whether the observed H$\alpha$ emission is likely due to chromospheric activity or to disk accretion, we plot our H$\alpha$ equivalent widths versus the adopted spectral type, and we compare with the empirical upper limit boundary of chromospheric activity derived in \citet{2003AJ....126.2997B}. The results are shown in Fig. \ref{fig4}. The 6 objects observed more than once are marked with red squares and the different measurements are connected with red lines. As mentioned in the previous section, the persistence of the strong H$\alpha$ emission indicates that it is not due to sporadic flares. All the objects occupy the region of the diagram above the chromospheric activity saturation line, and hence they are likely accreting matter from disks.

\begin{figure}%[h]
   \centering
%   \resizebox{\hsize}{!}{\includegraphics[width=0.8\textwidth, angle=270]{NaDprofinal.ps}}
   \resizebox{\hsize}{!}{\includegraphics[width=\textwidth]{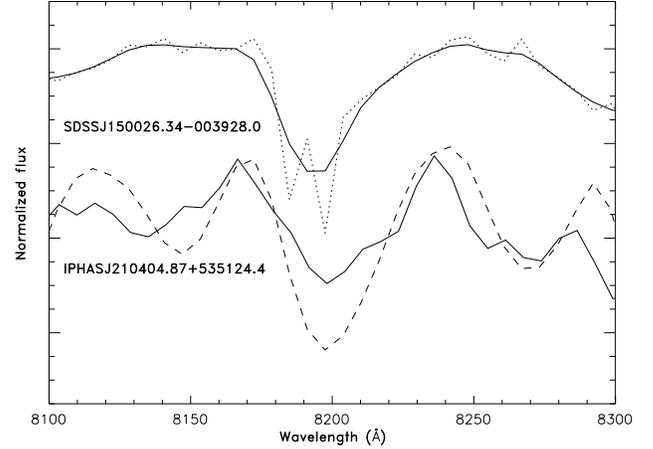}}
   \caption{Effects of the spectral resolution and residual fringing on our M6 template (SDSSJ150026.34-003928.0). \emph{Dotted line:} original spectrum of the template; \emph{solid line:} the same spectrum degraded to our spectral resolution; \emph{dashed line:} the same spectrum degraded to our spectral resolution, and with residual fringing. For comparison we also show the spectrum of a typical M6 object from our sample (IPHAS J210404.87+535124.4).} %Despite of the fringing residuals the doublet is clearly identified in the SDSSJ150026.34-003928.0.}}
   \label{fig5}
\end{figure}

\begin{figure}%[h]
   \centering
   \resizebox{\hsize}{!}{\includegraphics[width=\textwidth, angle=270]{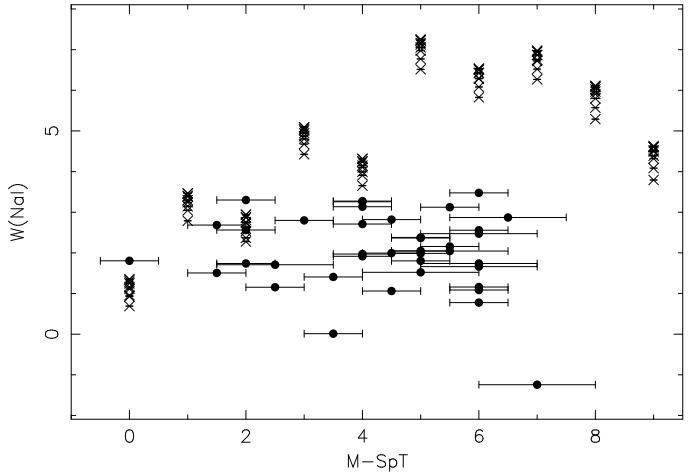}}
   \caption{Dependence of the \emph{W}(Na I) on the residual fringing and the spectral type. Filled circles denote the IPHAS objects and crosses denote measurements with different fringing residuals for the M objects used as reference to classify the IPHAS objects.}
   \label{fig5a}
   \end{figure}

\subsection{Surface gravity}\label{secsodio}

The NaI subordinate doublet at 818.3nm and 819.5nm is a good gravity indicator for late-M spectral types \citep{1991ApJS...77..417K, 1996ApJ...469..706M, 1998AJ....115.2074B}. This doublet becomes weaker as the surface gravity decreases.  We have checked whether the doublet can be identified on the reference objects despite of the fringing and low resolution of our targets. In Fig. \ref{fig5} we compare the SDSSJ150026.34-003928.0 reference dwarf (solid and dotted top lines and dashed bottom line) with one of our M6 classified objects IPHAS J210404.87+535124.4 (solid bottom line) and the doublet can be clearly be identified despite the residual fringing and low resolution. In Fig. \ref{fig5a} we show the results of the measurements described in \ref{ew}. We find that for M3 type and later the doublet can be clearly identified but not for earlier types since the doublet is weaker for this spectral types and is embedded on the residual fringing. According these results we can conclude that in all of our objects classified as M3-4 the NaI doublet is weaker in our H$\alpha$ emission objects than in field dwarfs of similar spectral class and much weaker for objects later than M4. This is consistent with our previous conclusion that most of our objects are very young and actively accreting. The half lifetime for disk accretion in VLM objects has been estimated to be less than 5 Myr by \citet{2005ApJ...626..498M}. According to this, the results suggest that our sample is $\leq$5 Myr.%, thus the results suggest that the NaI doublet is much weaker in our H$\alpha$ emission objects than in field dwarfs of similar spectral class, and thus to be very young objects.} %worse signal to noise ratio. Thus for those we could measure, the results must be taken as an upper limit}.   
%In Fig. \ref{fig5} we display a comparison of the NaI equivalent widths of our H$\alpha$ emission objects (open circles) with the NaI equivalent widths published by \citet{2004AJ....127..449M} in a spectroscopic study of VLM stars and BDs in the OB Upper Scorpius (USco) association (crosses). We also plot the NaI values obtained by us in the template dwarf spectra used in the  $\chi$$^2$ method (open triangles). 

%We find that the NaI doublet is much weaker in our H$\alpha$ emission objects than in field dwarfs of similar spectral class, and slightly weaker than among the VLM stars and BDs members in USco. The age of USco has been estimated at 5 Myr, and thus our results suggest that our sample is younger than 5 Myr. This is consistent with our previous conclusion that most of our objects are actively accreting, because the half lifetime for disk accretion in VLM objects has been estimated to be less than 5 Myr by {\bf \citet{2005ApJ...626..498M}}. \\

\begin{table*}
\caption{Photometric data of the IPHAS objects.}             
\label{table:2}      
\centering          
 \begin{tabular}{c c c c c c c c }     % 8 columns 
\hline\hline       
\scriptsize{IPHAS name/position} & \scriptsize{Observation} & \multicolumn{3}{ c }{\scriptsize{IPHAS magnitudes}} & \multicolumn{3}{ c }{\scriptsize{2MASS magnitudes}}\\
\scriptsize{J[RA(2000)+Dec.(2000)]}  & \scriptsize{date} & \scriptsize{r'} &\scriptsize{ r'-i'}& \scriptsize{r'- H$\alpha$} &\scriptsize{J}&\scriptsize{H}&\scriptsize{K} \\ 
\hline                    
\scriptsize{J001649.57+654241.8} &\scriptsize{2004-08-07} &\scriptsize{20.308$\pm$0.045} & \scriptsize{2.015$\pm$0.054} & \scriptsize{2.107$\pm$0.050} &\scriptsize{ 15.005$\pm$0.053}&\scriptsize{ 13.758$\pm$0.052} &\scriptsize{ 12.981$\pm$0.041} \\  
\scriptsize{J001655.91+654732.8} &\scriptsize{2004-08-07} &\scriptsize{18.803$\pm$0.014} & \scriptsize{2.209$\pm$0.017} & \scriptsize{1.593$\pm$0.018} &\scriptsize{ 13.645$\pm$0.026}& \scriptsize{12.506$\pm$0.023} & \scriptsize{11.993$\pm$0.022} \\  
\scriptsize{J011443.00+620820.9} &\scriptsize{2003-10-10} &\scriptsize{20.033$\pm$0.148} & \scriptsize{1.756$\pm$0.155} & \scriptsize{2.115$\pm$0.153} &\scriptsize{ 15.487$\pm$0.064}&\scriptsize{ 14.198$\pm$0.070} &\scriptsize{ 13.411$\pm$0.042} \\  
\scriptsize{J012348.67+614931.8} &\scriptsize{2003-10-12} &\scriptsize{19.716$\pm$0.081} & \scriptsize{1.722$\pm$0.086} & \scriptsize{2.102$\pm$0.083} &\scriptsize{ 15.636$\pm$0.061}&\scriptsize{ 14.541$\pm$0.060} &\scriptsize{ 14.039$\pm$0.053} \\  
\scriptsize{J013720.01+645957.7} &\scriptsize{2003-10-14} &\scriptsize{18.001$\pm$0.028} & \scriptsize{1.721$\pm$0.030} & \scriptsize{1.318$\pm$0.031} &\scriptsize{ 13.621$\pm$0.028}&\scriptsize{ 12.687$\pm$0.036} &\scriptsize{ 12.083$\pm$0.023} \\  
\scriptsize{J023616.00+615609.6} &\scriptsize{2003-11-10} &\scriptsize{19.727$\pm$0.087} & \scriptsize{1.280$\pm$0.096} & \scriptsize{2.041$\pm$0.090} &\scriptsize{ 15.984$\pm$0.099}&\scriptsize{ 15.022$\pm$0.103} &\scriptsize{ 14.285$\pm$0.076} \\  
\scriptsize{J035449.17+530903.3} &\scriptsize{2004-12-01} &\scriptsize{20.249$\pm$0.059} & \scriptsize{1.961$\pm$0.071} & \scriptsize{1.906$\pm$0.072} &\scriptsize{ 15.719$\pm$0.087}&\scriptsize{ 14.719$\pm$0.088} &\scriptsize{ 14.055$\pm$0.072} \\  
\scriptsize{J035823.95+522312.6} &\scriptsize{2004-12-29} &\scriptsize{19.016$\pm$0.038} & \scriptsize{1.283$\pm$0.057} & \scriptsize{1.965$\pm$0.042} &\scriptsize{ 15.745$\pm$0.061}&\scriptsize{ 14.826$\pm$0.061} &\scriptsize{ 14.298$\pm$0.064} \\  
\scriptsize{J042450.68+455330.2} &\scriptsize{2004-10-22} &\scriptsize{18.619$\pm$0.017} & \scriptsize{1.732$\pm$0.021} & \scriptsize{1.662$\pm$0.021} &\scriptsize{ 14.116$\pm$0.030}&\scriptsize{ 12.76 $\pm$0.026} &\scriptsize{ 11.849$\pm$0.022} \\  
\scriptsize{J183034.15+003800.6} &\scriptsize{2004-06-08} &\scriptsize{18.368$\pm$0.010} & \scriptsize{1.907$\pm$0.013} & \scriptsize{1.983$\pm$0.012} &\scriptsize{ 13.633$\pm$0.028}&\scriptsize{ 12.524$\pm$0.031} &\scriptsize{ 11.937$\pm$0.024} \\  
\scriptsize{J183753.25+001849.2} &\scriptsize{2004-06-09} &\scriptsize{19.134$\pm$0.018} & \scriptsize{2.621$\pm$0.021} & \scriptsize{1.566$\pm$0.024} &\scriptsize{ 13.05 $\pm$0.026}&\scriptsize{ 11.837$\pm$0.021} &\scriptsize{ 11.16 $\pm$0.022} \\  
\scriptsize{J192656.04+211438.0} &\scriptsize{2004-08-06} &\scriptsize{16.667$\pm$0.005} & \scriptsize{1.195$\pm$0.008} & \scriptsize{1.313$\pm$0.007} &\scriptsize{ 12.347$\pm$0.022}&\scriptsize{ 11.39 $\pm$0.021} &\scriptsize{ 10.742$\pm$0.019} \\  
\scriptsize{J202050.40+394243.6} &\scriptsize{2007-06-21} &\scriptsize{19.954$\pm$0.037} & \scriptsize{1.110$\pm$0.056} & \scriptsize{1.757$\pm$0.049} &\scriptsize{ 15.83 $\pm$0.076}&\scriptsize{ 14.569$\pm$0.051} &\scriptsize{ 13.611$\pm$0.050} \\  
\scriptsize{J202434.30+422126.4} &\scriptsize{2003-08-10} &\scriptsize{18.304$\pm$0.040} & \scriptsize{1.912$\pm$0.042} & \scriptsize{1.170$\pm$0.046} &\scriptsize{ 13.335$\pm$0.024}&\scriptsize{ 12.249$\pm$0.027} &\scriptsize{ 11.501$\pm$0.023} \\  
\scriptsize{J202437.28+385806.9} &\scriptsize{2006-11-04} &\scriptsize{19.879$\pm$0.070} & \scriptsize{1.921$\pm$0.078} & \scriptsize{1.664$\pm$0.080} &\scriptsize{ 15.337$\pm$0.055}&\scriptsize{ 14.293$\pm$0.052} &\scriptsize{ 13.685$\pm$0.055} \\  
\scriptsize{J202455.53+424504.0} &\scriptsize{2004-08-08} &\scriptsize{18.902$\pm$0.014} & \scriptsize{1.910$\pm$0.018} & \scriptsize{1.525$\pm$0.020} &\scriptsize{ 14.425$\pm$0.037}&\scriptsize{ 13.548$\pm$0.038} &\scriptsize{ 12.883$\pm$0.030} \\  
\scriptsize{J202759.82+390418.2} &\scriptsize{2006-11-24} &\scriptsize{17.947$\pm$0.013} & \scriptsize{1.450$\pm$0.017} & \scriptsize{1.820$\pm$0.015} &\scriptsize{ 13.729$\pm$0.030}&\scriptsize{ 12.603$\pm$0.026} &\scriptsize{ 11.897$\pm$0.022} \\  
\scriptsize{J204218.54+395723.8} &\scriptsize{2007-06-29} &\scriptsize{17.712$\pm$0.009} & \scriptsize{1.468$\pm$0.012} & \scriptsize{0.729$\pm$0.015} &\scriptsize{ 14.152$\pm$0.028}&\scriptsize{ 13.099$\pm$0.029} &\scriptsize{ 12.587$\pm$0.026} \\  
\scriptsize{J204350.68+400108.8} &\scriptsize{2007-06-29} &\scriptsize{17.273$\pm$0.007} & \scriptsize{1.270$\pm$0.010} & \scriptsize{1.223$\pm$0.010} &\scriptsize{ 14.035$\pm$0.031}&\scriptsize{ 13.024$\pm$0.023} &\scriptsize{ 12.44 $\pm$0.026} \\  
\scriptsize{J204704.82+434911.4} &\scriptsize{2003-10-12} &\scriptsize{18.758$\pm$0.032} & \scriptsize{1.232$\pm$0.037} & \scriptsize{2.049$\pm$0.034} &\scriptsize{ 15.336$\pm$0.054}&\scriptsize{ 14.343$\pm$0.051} &\scriptsize{ 13.764$\pm$0.043} \\  
\scriptsize{J205613.08+443424.2} &\scriptsize{2003-11-10} &\scriptsize{20.250$\pm$0.113} & \scriptsize{1.958$\pm$0.121} & \scriptsize{2.267$\pm$0.120} &\scriptsize{ 14.678$\pm$0.043}&\scriptsize{ 13.895$\pm$0.056} &\scriptsize{ 13.387$\pm$0.045} \\  
\scriptsize{J205701.63+434138.7} &\scriptsize{2007-06-23} &\scriptsize{18.764$\pm$0.013} & \scriptsize{2.269$\pm$0.016} & \scriptsize{1.373$\pm$0.019} &\scriptsize{ 13.952$\pm$0.029}&\scriptsize{ 13.142$\pm$0.032} &\scriptsize{ 12.649$\pm$0.029} \\  
\scriptsize{J205702.69+434143.6} &\scriptsize{2007-06-23} &\scriptsize{19.871$\pm$0.029} & \scriptsize{2.252$\pm$0.034} & \scriptsize{1.868$\pm$0.036} &\scriptsize{ 14.348$\pm$0.055}&\scriptsize{ 13.382$\pm$0.041} &\scriptsize{ 12.695$\pm$0.039} \\  
\scriptsize{J210404.87+535124.4} &\scriptsize{2006-10-02} &\scriptsize{18.299$\pm$0.015} & \scriptsize{1.464$\pm$0.020} & \scriptsize{2.507$\pm$0.016} &\scriptsize{ 13.432$\pm$0.025}&\scriptsize{ 12.401$\pm$0.025} &\scriptsize{ 11.613$\pm$0.025} \\  
\scriptsize{J213528.41+575823.0} &\scriptsize{2004-08-05} &\scriptsize{18.640$\pm$0.015} & \scriptsize{1.360$\pm$0.025} & \scriptsize{1.980$\pm$0.019} &\scriptsize{ 15.15 $\pm$0.059}&\scriptsize{ 14.212$\pm$0.056} &\scriptsize{ 13.598$\pm$0.044} \\ 
\scriptsize{J213545.87+573640.1} &\scriptsize{2004-08-05} &\scriptsize{18.526$\pm$0.014} & \scriptsize{1.509$\pm$0.021} & \scriptsize{1.996$\pm$0.017} &\scriptsize{ 14.551$\pm$0.037}&\scriptsize{ 13.676$\pm$0.034} &\scriptsize{ 13.143$\pm$0.033} \\ 
\scriptsize{J213938.83+575451.4} &\scriptsize{2004-08-27} &\scriptsize{19.637$\pm$0.044} & \scriptsize{1.658$\pm$0.053} & \scriptsize{2.083$\pm$0.049} &\scriptsize{ 14.757$\pm$0.039}&\scriptsize{ 13.621$\pm$0.040} &\scriptsize{ 13.032$\pm$0.033} \\  
\scriptsize{J214547.73+564845.7} &\scriptsize{2004-08-28} &\scriptsize{18.942$\pm$0.035} & \scriptsize{2.058$\pm$0.039} & \scriptsize{1.740$\pm$0.040} &\scriptsize{ 14.333$\pm$0.032}&\scriptsize{ 13.399$\pm$0.036} &\scriptsize{ 12.822$\pm$0.025} \\  
\scriptsize{J214625.99+572829.0} &\scriptsize{2004-08-29} &\scriptsize{18.592$\pm$0.036} & \scriptsize{1.377$\pm$0.046} & \scriptsize{2.315$\pm$0.038} &\scriptsize{ 14.069$\pm$0.034}&\scriptsize{ 13.064$\pm$0.033} &\scriptsize{ 12.301$\pm$0.026} \\  
\scriptsize{J222025.22+605423.7} &\scriptsize{2003-11-03} &\scriptsize{19.167$\pm$0.051} & \scriptsize{1.459$\pm$0.057} & \scriptsize{2.130$\pm$0.053} &\scriptsize{ 15.32 $\pm$0.053}&\scriptsize{ 14.413$\pm$0.051} &\scriptsize{ 13.911$\pm$0.052} \\  
\scriptsize{J224830.78+611417.9} &\scriptsize{2004-08-05} &\scriptsize{19.650$\pm$0.034} & \scriptsize{2.069$\pm$0.040} & \scriptsize{1.988$\pm$0.040} &\scriptsize{ 14.959$\pm$0.043}&\scriptsize{ 13.967$\pm$0.040} &\scriptsize{ 13.458$\pm$0.036} \\  
\scriptsize{J225331.29+623543.3} &\scriptsize{2004-08-21} &\scriptsize{20.554$\pm$0.051} & \scriptsize{2.153$\pm$0.062} & \scriptsize{1.997$\pm$0.060} &\scriptsize{ 15.393$\pm$0.059}&\scriptsize{  14.17$\pm$0.046} &\scriptsize{ 13.471$\pm$0.035} \\  
\scriptsize{J225923.71+614138.6} &\scriptsize{2007-06-27} &\scriptsize{19.228$\pm$0.016} & \scriptsize{1.493$\pm$0.023} & \scriptsize{2.219$\pm$0.019} &\scriptsize{ 15.341$\pm$0.071}&\scriptsize{ 14.192$\pm$0.057} &\scriptsize{ 13.487$\pm$0.051} \\    
\hline                  
\end{tabular}
\end{table*}

\begin{figure*}
\begin{tabular}{cc}
\centering
  \resizebox{0.5\hsize}{!}{\includegraphics[width=\textwidth]{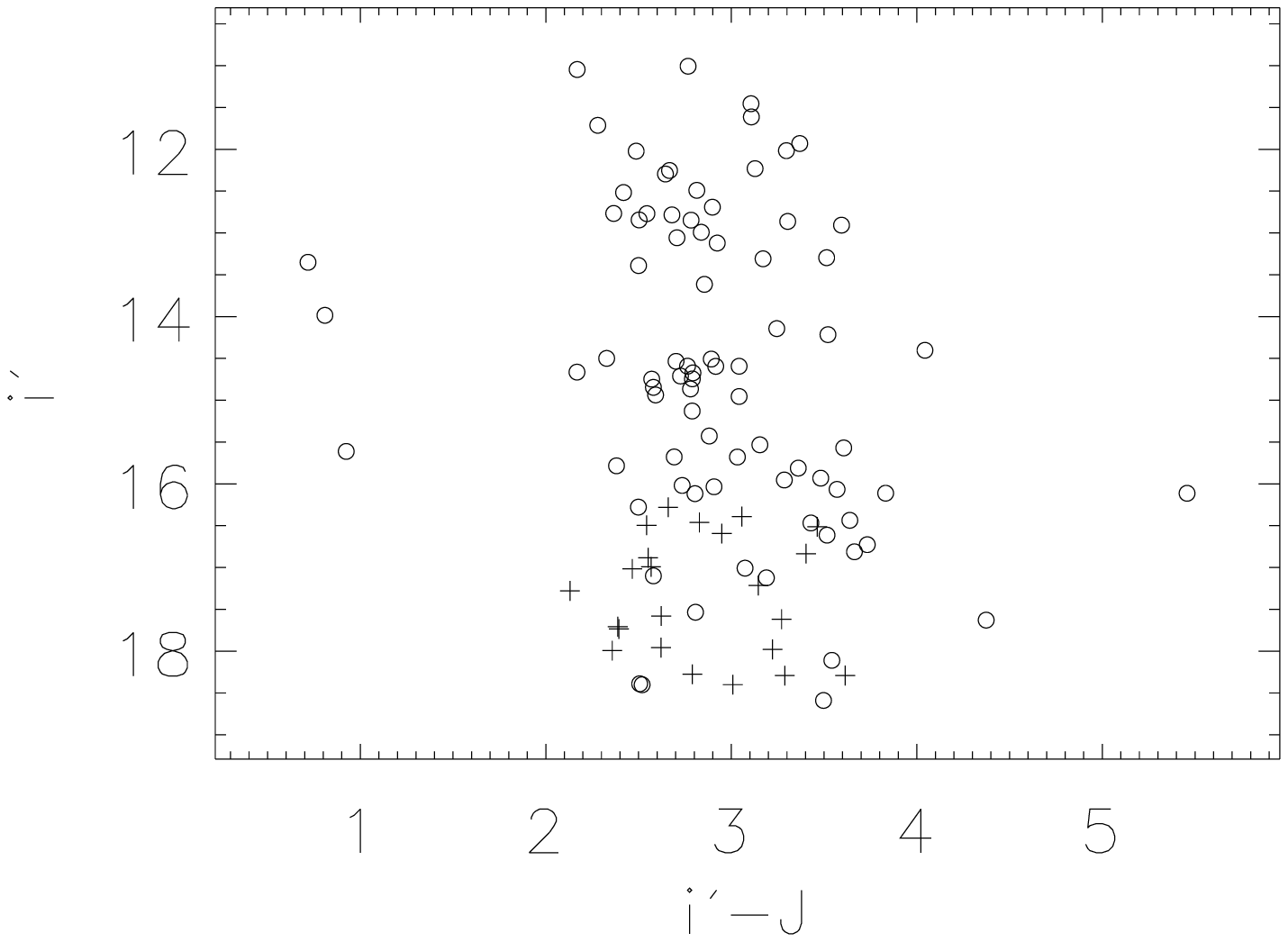}}
  \resizebox{0.5\hsize}{!}{\includegraphics[width=\textwidth]{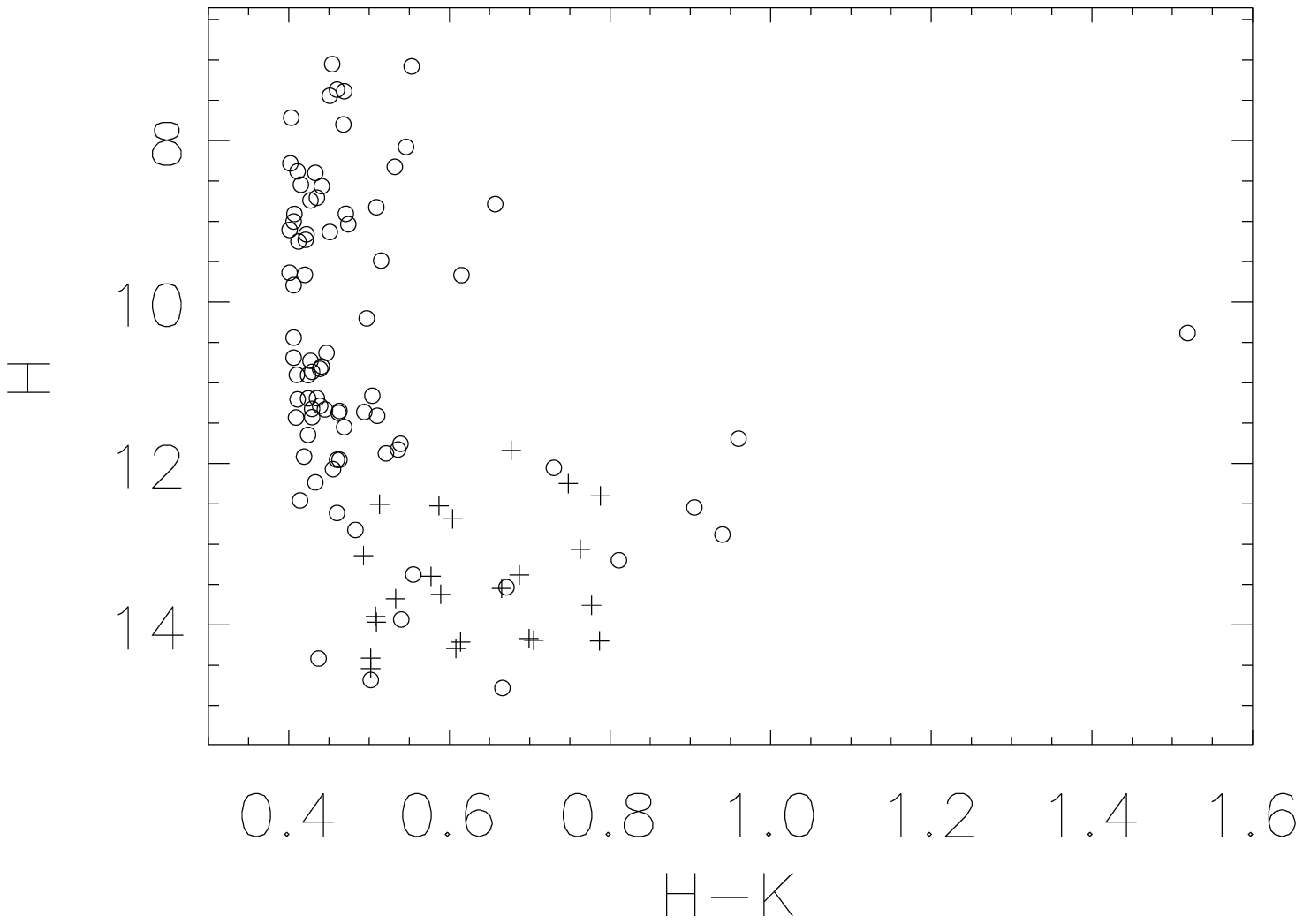}}
\end{tabular}
      \caption{ Colour-magnitude diagrams of the candidates. Open circles represent the objects observed spectroscopically; plus signs identify the observed objects that show H$\alpha$ emission and are classified as M4 or later.} 
         \label{fig6}
\end{figure*}

\begin{figure*}
\centering
\begin{tabular}{cc}
  \resizebox{0.5\hsize}{!}{\includegraphics[width=\textwidth]{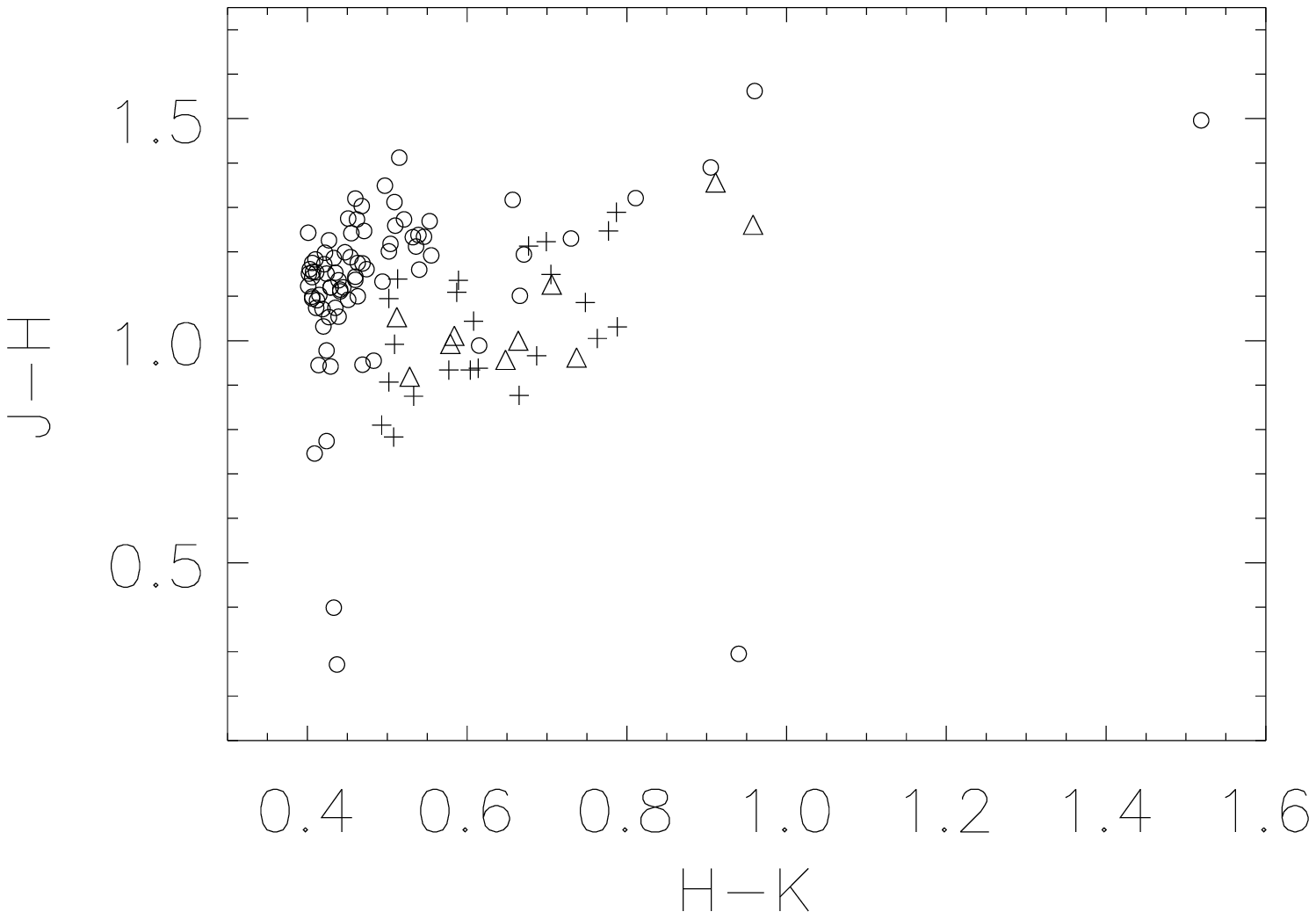}}
  \resizebox{0.5\hsize}{!}{\includegraphics[width=\textwidth]{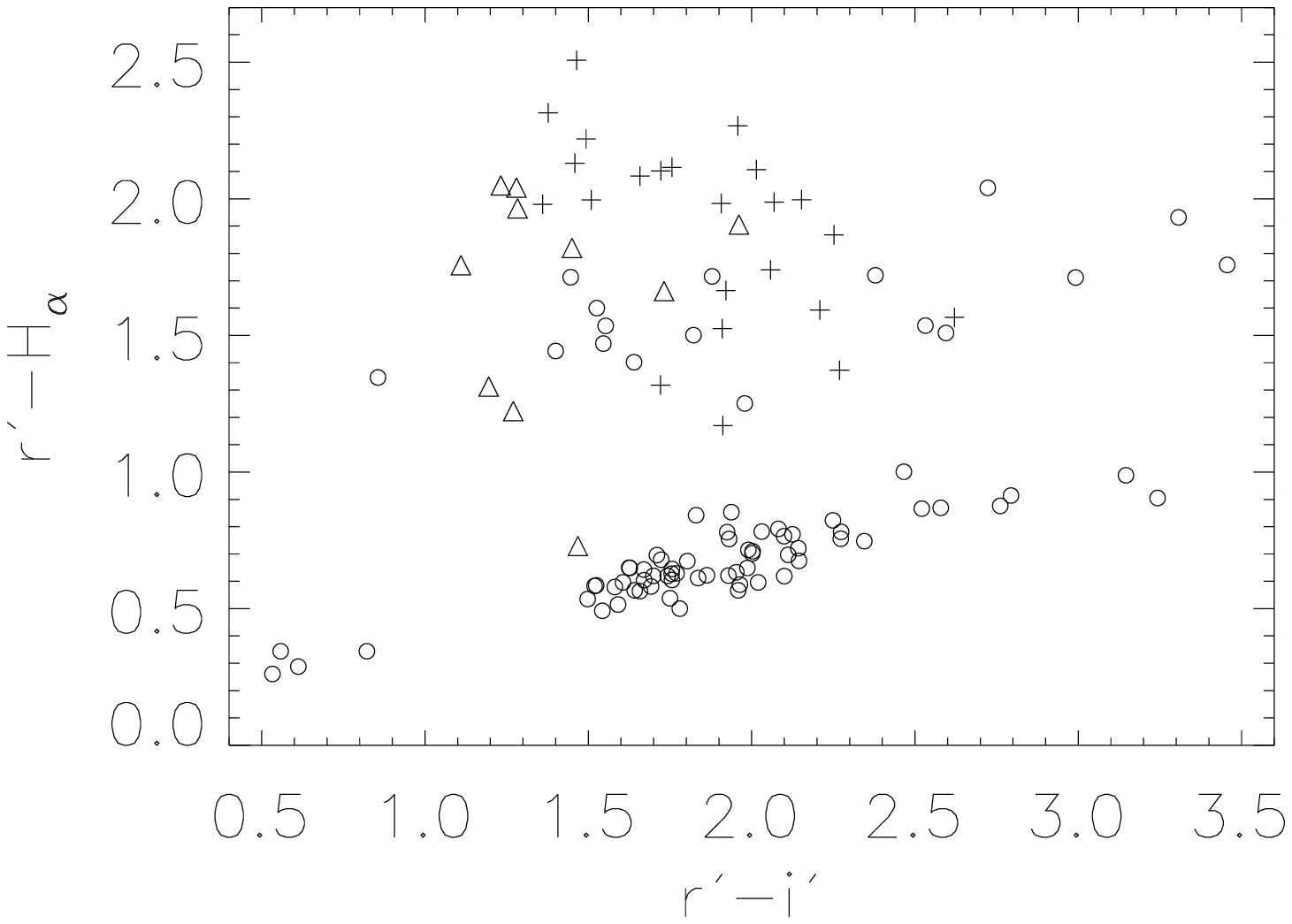}}
\end{tabular}
      \caption{ Colour-colour diagrams. Open circles represent all objects observed 
spectroscopically; open triangles represent the objects that show H$\alpha$ emission and 
are classified as earlier than M4, and plus signs represent the objects that show H$\alpha$  emission and are classified as M4 or later.} 
         \label{fig7}
 \end{figure*}
 \begin{figure*}
	\resizebox{0.5\hsize}{!}{\includegraphics[width=\textwidth,angle=270]{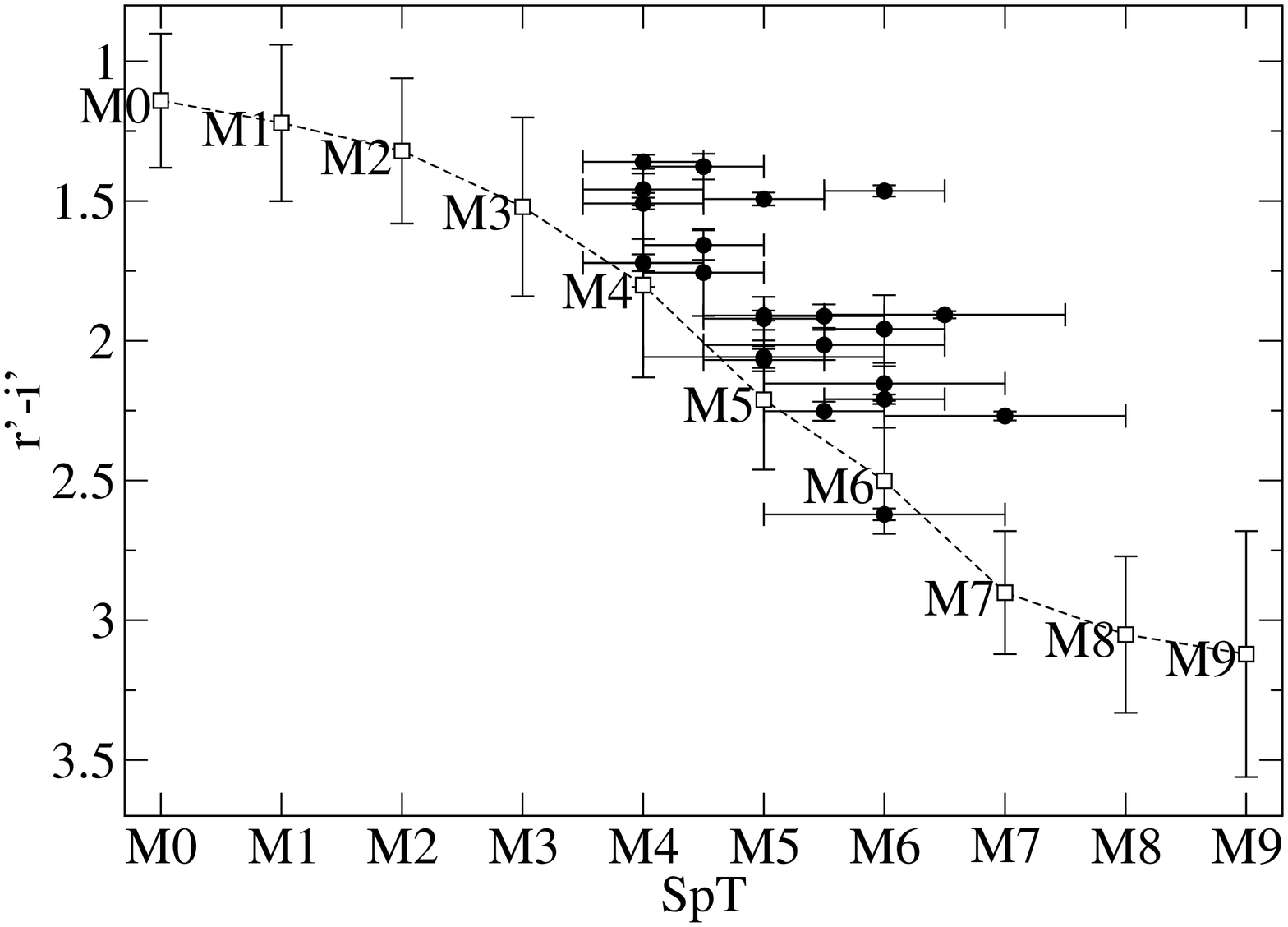}}
        \resizebox{0.5\hsize}{!}{\includegraphics[width=\textwidth,angle=270]{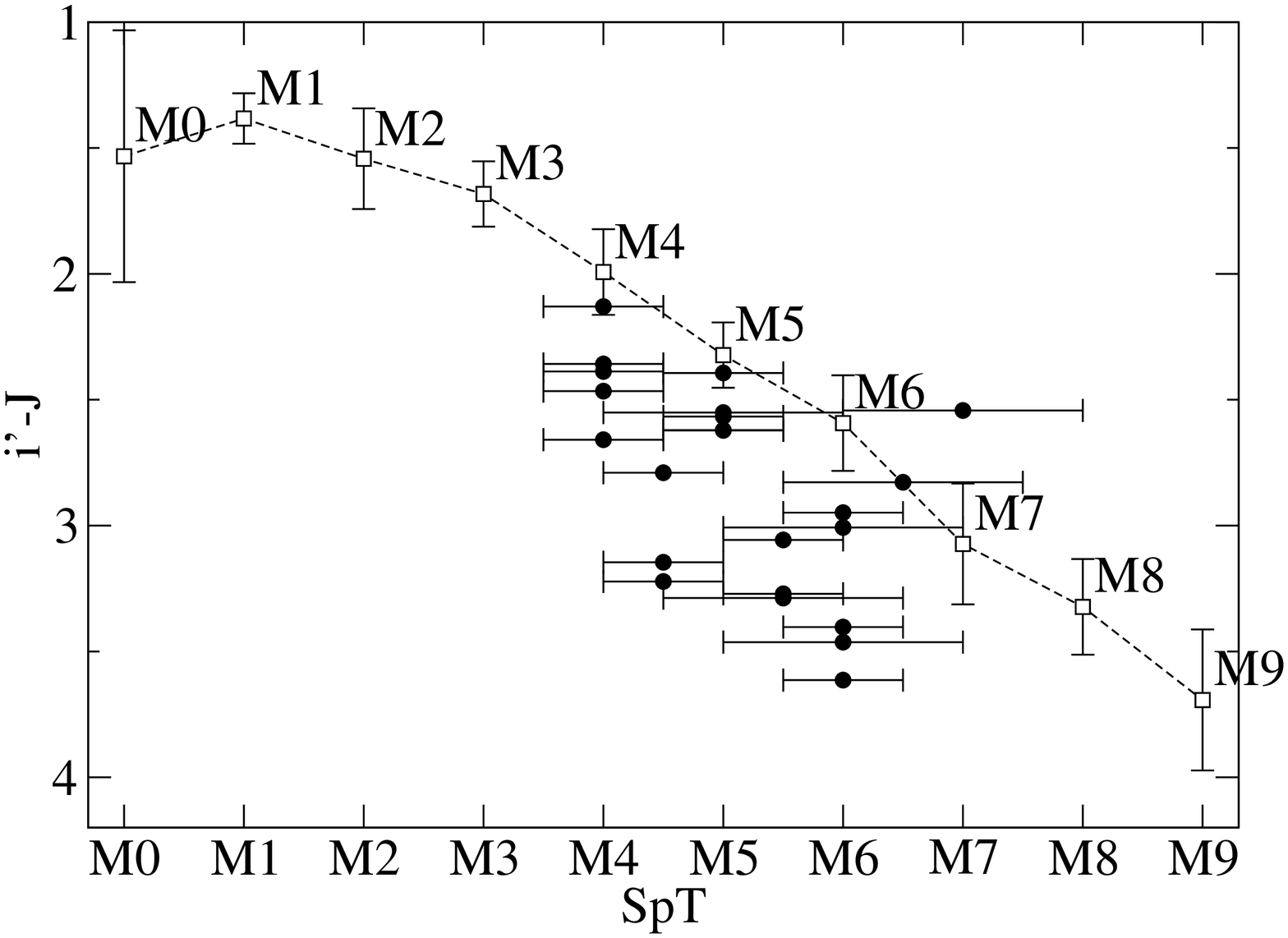}}	
        \caption{Comparison of the results for the IPHAS objects (filled circles) and the characterisation for M dwarfs of \citet{2002AJ....123.3409H} (open squares).}
\label{fig8}
\end{figure*}

\subsection{Photometric properties}
The main emphasis of our analysis has been on the spectroscopic properties of our sample. 
However, it is also interesting to study the photometric properties. 
The photometry according to best new IPHAS photometry of our 33 main objects is given in Table \ref{table:2}. colour-magnitude (Fig. \ref{fig6}) and colour-colour (Fig. \ref{fig7}) diagrams are used to check if our sample of 33 interesting objects stand out from the crowd.\\ 
In Fig.\ref{fig6} we have represented with a plus sign the objects classified as M4 or later. It is observed that they tend to populate the fainter part of the diagram. All of them have magnitudes of r'$>$ 16.5, J $>$12, i' $>$ 15.5 and H $>$ 11.\\ 
In Fig. \ref{fig7}, we plot also the objects that show emission in H$\alpha$ but are classified as earlier than M4 as open triangles and those classified as M4 or later with plus sign. All the accretors show red colours with H$-$K $>$ 0.5, and most of them  i'$-$J $<$ 4. We find that the r'$-$i' \ and r'$-$H colours tend to the blue, which could be related to the 
presence of optical veiling in the r' band. On the other hand, the red H$-$K colours could be ascribed to warm dust emission.

 We have compared the colour-spectral type relation for our targets classified as M4 or later with those in the study of \citet{2002AJ....123.3409H} in which the spectroscopic characterisation of SDSS M dwarfs was done (see Fig. \ref{fig8}). In general our objects follow the relation observed in the SDSS objects, but there are some significant discrepancies. In the r'$-$i' \ vs. SpT diagram, it is observed a tendency to the blue of some objects compared with SDSS locus. 
On the other hand, for the colour i'$-$J, it is observed a tendency to the red for our sample. 
These discrepancies could be due to the effects of accretion and low surface gravity on the spectral energy distribution of VLM objects. 

\begin{acknowledgements}
Part of the data presented here have been taken using ALFOSC, which is owned by the Instituto de Astrof\'isica de Andaluc\'ia (IAA) and operated at the Nordic Optical Telescope under agreement between IAA and the NBIfAFG of the Astronomical Observatory of Copenhagen. This paper makes use of data obtained as part of the INT Photometric H$\alpha$ Survey of the Northern Galactic Plane (IPHAS) carried out at the Isaac Newton Telescope. The INT and WHT are operated on the island of La Palma by the Isaac Newton Group in the Spanish Observatorio del Roque de los Muchachos of the Instituto de Astrof\'isica de Canarias. All IPHAS data are processed by the Cambridge Astronomical Survey Unit, at the Institute of Astronomy in Cambridge. This publication makes use of data products from the Two Micron All Sky Survey, which is a joint project of the University of Massachusetts. This work makes use of EURO-VO software, tools or services. The EURO-VO has been funded by the European Commission through contract numbers RI031675 (DCA) and 011892 (VO-TECH) under the 6th Framework Programme. This research has made use of the Spanish Virtual Observatory supported from the Spanish MEC through grants AyA2005-04286, AyA2005-24102-E. This project has also been supported by MEC, through grant AyA2007-67458.
\end{acknowledgements}

%############################################################
%	BIBLIOGRAPHY
%############################################################
\bibliographystyle{aa}

\end{document}